\documentclass[proceedings]{stacs}
\stacsheading{2009}{183--194}{Freiburg}
\firstpageno{183}

\usepackage{amsfonts, amsmath, amssymb}
\usepackage[T1]{fontenc} %
\usepackage[latin1]{inputenc} %
\def\MIT{M{\footnotesize{ULTICUT}} I{\footnotesize{N}} T{\footnotesize{REES}} }
\def\MITB{M{\footnotesize{ULTICUT}} I{\footnotesize{N}} T{\footnotesize{REES}}}
\def\MIC{M{\footnotesize{ULTICUT}} I{\footnotesize{N}} C{\footnotesize{ATERPILLARS}} }

\newtheorem{claim}{Claim}

\title{A polynomial kernel for Multicut In Trees}

\author[lab1]{N. Bousquet}{Nicolas Bousquet}
\address[lab1]{ENS Cachan, 61, avenue du Pr\'{e}sident Wilson, 94235 Cachan cedex - France}
\email{nbousque@dptinfo.ens-cachan.fr}
\author[lab2]{J. Daligault}{Jean Daligault}
\address[lab2]{Universit\'{e} Montpellier II - CNRS, LIRMM, 161 rue Ada
34392 Montpellier Cedex 5 - France}\email{daligault@lirmm.fr} 
\author[lab2]{S. Thomass\'e}{St\'ephan Thomass\'e}\email{thomasse@lirmm.fr} 
\author[lab3]{A. Yeo}{\\Anders Yeo}
\address[lab3]{Royal Holloway, University of London, Egham Hill, EGHAM,
TW20 0EX - UK}\email{anders@cs.rhul.ac.uk}
\thanks{Part of this research was supported by Alliance Project "Partitions de graphes orient\'{e}s". Part of this research was supported by ANR Project GRAAL}

\begin{document}
\maketitle

\begin{abstract}
\par The \MIT problem consists in deciding, given a tree, a set of requests (i.e. paths in the tree) and an integer $k$, whether there exists a set of $k$ edges cutting all the requests. This problem was shown to be FPT by Guo and Niedermeyer in \cite{GuoNied}. They also provided an exponential kernel. They asked whether this problem has a polynomial kernel. This question was also raised by Fellows in~\cite{MF}.
\par We show that \MIT has a polynomial kernel.
\end{abstract}

\section{Introduction}
An efficient way of dealing with NP-hard problems
is to identify a parameter which contains its 
computational hardness. For instance, instead 
of asking for a minimum vertex cover in a graph -
a classical NP-hard optimization question - one
can ask for an algorithm which would decide, in
$O(f(k).n^d)$ time for some fixed $d$, if a graph 
of size $n$ has a  vertex cover of size at 
most $k$. If such an algorithm exists, the problem
is called {\it fixed-parameter tractable}, or FPT for 
short. An extensive litterature is devoted to FPT, 
the reader is invited to read \cite{DF2}, \cite{FL}
and \cite{RN}.

Kernelization is a natural way of proving that a 
problem is FPT. Formally, a {\it kernelization algorithm}
receives as input an instance $(I,k)$ of the parameterized
problem, and outputs, in polynomial time  in the size of 
the instance, another instance $(I',k')$ such that

\begin{itemize}
\item $k'\leq k$, 
\item the size of $I'$ only depends of $k$, 
\item the instances $(I,k)$ and $(I',k')$ are both true or both 
false. 
\end{itemize}

The reduced instance $(I',k')$ is called a {\it kernel}.
The existence of a kernelization algorithm 
clearly implies the FPT character of the problem since 
one can kernelize the instance, and then solve the reduced
instance $G',k'$ using brute force, hence giving an 
$O(f(k)+n^d)$ algorithm. 
A classical result asserts that being FPT is indeed equivalent to 
having kernelization. The drawback of this result is that 
the size of the reduced instance $G'$ is not necessarily 
small with respect to $k$. A much more constrained condition
is to be able to reduce to an instance of polynomial 
size in terms of $k$. Consequently, in the zoology of parameterized 
problems, the first distinction is done between three classes: 
W[1]-hard, FPT, polykernel. 

A kernelization algorithm can be used as a preprocessing step to reduce the size of the instance before applying an algorithm. Being able to ensure that this kernel has actually polynomial size in $k$ enhances the overall speed of the algorithm. See \cite{GuoNied2} for a recent review on kernalization.

The existence of a polynomial kernel can be a subtle issue. A recent result by Fernau et al \cite{fernau} shows that Rooted $k$-Leaf Outbranching has a cubic kernel while $k$-Leaf Outbranching does not, unless polynomial hierarchy collapses to third level, using a breakthrough lower bound result by Bodlaender and al \cite{NoKernel}.

In the (unweighted) \MIT problem, we consider a tree $T$ together with a set $P$ 
of pairs of distinct nodes of $T$, called \emph{requests}. Hence, a request 
can also be seen as a prescribed path joining these two nodes.
We will often identify the request and its path.
A \emph{multicut} of $(T,P)$ is a set $S$ of edges of $T$ which intersect 
every request in $P$, i.e. every path corresponding to a request contains
an edge of $S$. 


\begin{problem}
\MITB: 

\vspace{12pt}
\textbf{Input}: A tree $T=(V,E)$, a set of requests $P$, an integer $k$.

\textbf{Output}: TRUE if there is a multicut of size at most $k$, otherwise FALSE.
\vspace{12pt}
\end{problem}



\par Note that a more general presentation of this problem is to assign weights to edges, and ask for a multicut of minimal weight. Our technique does not seem to generalize to the weighted case.
\par This problem appears in network issues (routing, telecommunication, ...). See \cite{survey} for a survey on multicommodity flow problems and multicut problems. It was shown in \cite{garg} that \MIT is NP-complete, and its associated decision problem is MaxSNP-hard and has a factor-2 polynomial time approximation algorithm.
\par This problem is known to be FPT, see~\cite{guothese} or~\cite{GuoNied} for a branching algorithm and 
an exponential kernel. The existence of a polynomial kernel was asked in~\cite{MF}. 
We verify that \MIT has indeed an $O(k^6)$ kernel. Our reduction is very much inspired 
from~\cite{guothese} and~\cite{GuoNied}.
In the next section, we first illustrate our techniques when the tree $T$
is a caterpillar. In Section 3 we extend the proof to general trees.

\section{A polynomial kernel for caterpillars}
A node of $T$ which is not a leaf is an \emph{internal node}. 
The \emph{internal tree} of $T$ is the tree restricted to its internal nodes.
We say that $T$ is a \emph{caterpillar} if its internal tree is a path. 
We consider the restriction of the \MIT problem to caterpillars, as 
it contains the core of our proof in the general case. 

Let us give some general definitions which will apply both for the caterpillar case and for the general case.

\par We say that two nodes $x$ and $y$ are \emph{R-neighbors} if there exists 
a request $xy$. A leaf $x$ and an internal node $y$ are \emph{quasi-R-neighbors} 
if there exists a request $xy$, or a request $xz$, where $z$ is a leaf rooted 
at $y$. An internal node with no leaf attached to it is an \emph{inner node}. If $x$ is a leaf, we denote by $e(x)$ and call \emph{the edge of $x$} the edge adjacent to $x$. A \emph{group} of leaves is the set of leaves connected to the same internal node. A \emph{group request} is a request $xy$ where $x$ and $y$ belong to the same group. A leaf which is an endpoint of a group request is a \emph{bad leaf}.
A \emph{leaf to leaf request} is a request between two leaves. An \emph{internal request} is a request between two internal nodes. A request between an internal node and a leaf is a \emph{mixed-request}. Two requests are \emph{disjoint} if their edge sets are disjoint. Two requests $x_{1}y_{1}$ and $x_{2}y_{2}$ are \emph{endpoint-disjoint} if $x_{1},y_{1},x_{2},y_{2}$ are pairwise different.
\par The \emph{internal path} of a request is the intersection between the path of the request and the internal tree. The \emph{common factor} of two requests is the intersection of their paths. A request $R_1$\emph{ dominates} a request $R_2$ if the internal path of $R_1$ contains the internal path of $R_2$.
\par \emph{Contracting} an edge $e$ in $(T,P)$ means contracting $e$ in $T$, and transforming each request of the form $(e_1,\dots,e_t,e,e_{t+1},\dots,e_l)$ in $P$ into $(e_1,\dots,e_t,e_{t+1},\dots,e_l)$. \emph{Deleting} an edge $e$ means contracting $e$ in $T$ and removing every request containing $e$ from $P$.

\par Two requests of length at least 2 from a given leaf $x$ \emph{have the same direction} if the second edge of their path starting at $x$ is the same. Two requests from an internal node $x$ \emph{have the same direction} if the first edge of their paths (starting at $x$) is the same. All the requests from $x$ \emph{have the same direction} if they pairwise have the same direction.

In the following, 
our instance $T$ is assumed to be a caterpillar. We call the two extremities of the internal path the
\emph{left end} and the \emph{right end} of $T$. The path between a node $x$ and the right (resp. left) end will be called \emph{right} and \emph{left} relatively to $x$.
\par Let $T'$ be the internal tree of the caterpillar $T$. The following five sets partition $T$:
\begin{itemize}
\item The set $I_1$ of leaves of $T'$.
\item The set $I_2$ of degree two nodes of $T'$.
\item The set $L_1$ of leaves rooted at $I_1$.
\item The set $L_2'$ of bad leaves rooted at $I_2$.
\item The set $L_2$ of the other leaves rooted at $I_2$.

\end{itemize}
\par The \emph{wingspan} $W$ of a leaf $x$ is the path between the closest quasi-R-neighbor on the right of $x$ and the closest quasi-R-neighbor on the left of $x$ (if no such neighbor exists, we take the father $f(x)$ of $x$ by convention). The \emph{size} of a wingspan is the number of $L_2$-leaves pending from it. The \emph{subcaterpillar} of the wingspan $W$ consists in $W$ and the leaves rooted at $W$. The wingspan $W$ \emph{dominates} a request $yz$ if both $y$ and $z$ belong to the subcaterpillar of $W$.
\par The usual way of exhibiting a kernel is to define a set of \emph{reduction rules}. These rules should be \emph{safe}, meaning that after applying a rule, the truth value of the problem on the instance does not change. Moreover the repeated application of the rules should take polynomial time. Finally, after iterating these rules on an instance, we want the \emph{reduced instance} to be of polynomial size in $k$.

\paragraph{\bf The reduction rules}
\par We apply the following reduction rules to an instance:
\begin{itemize}
\item[(0)] Unit Request: if a request $R$ has length one, i.e. $R=e$ for some edge $e$ of $T$, then we delete $e$ and decrease $k$ by one.
\item[(1)] Disjoint Requests: if there are $k+1$ disjoint requests in $P$, then we return a trivially false instance.
\item[(2)] Unique Direction: if all the requests starting at a leaf $x$ have the same direction, then contract $e(x)$. If all the requests starting at an inner node $x$ have the same direction, then contract the edge $e$ adjacent to $x$ which does not belong to any request starting at $x$.
\item[(3)] Inclusion: if a request $R$ is included in another request $R'$, then delete $R'$ from the set of requests. 
\item[(4)] Common Factor: let $R$ be a request. If $k+1$ requests $R_1,\dots,R_{k+1}$ different from $R$ but intersecting $R$ are such that for every $i\neq j$, the common factor of $R_i$ and $R_j$ is a subset of $R$, then delete $R$ from the set of requests.
\item[(5)] Dominating Wingspan: if $x$ is an $L_2$-leaf with a wingspan dominating at least $k+1$ endpoint-disjoint leaf to leaf or mixed requests, then contract $e(x)$. 
\end{itemize}
\par Each iteration of the reduction consists in applying the first applicable rule, in the above order.

\begin{lemma}
Rules Unit Request, Disjoint Requests, Unique Direction, Inclusion, Common Factor and Dominating Wingspan are safe.
\end{lemma}
\proof
\begin{itemize}
\item[(0)] Rule Unit Request is obvious.
\item[(1)] Rule Disjoint Requests is obvious.
\item[(2)] For Rule Unique Direction, assume first that all the requests from a leaf $x$ have the same direction, and that a multicut contains $e(x)$. Let $e'$ be the second common edge of all these paths. As $e'$ cuts all the requests cut by $e(x)$, if $e(x)$ is in a solution $S$ then $S \backslash \{e(x)\} \cup \{e'\}$ is also a solution. So we can contract $e(x)$. 
Now, assume that all the requests from an inner node $x$ go to the right. If a solution $S$ contains the edge $e$ adjacent to $x$ on the left then $S\backslash \{e\} \cup \{e'\}$, where $e'$ is the right edge adjacent to $x$, is a solution since a request going through $e$ also goes through $e'$.
\item[(3)] For Rule Inclusion, observe that an edge cutting $R$ also cuts all the paths containing $R$.
\item[(4)] If there is a multicut of $k$ edges, then one of these edges must intersect two requests among the $k+1$ mentioned in Rule Common Factor. This edge lies in the intersection of two paths, hence in $R$, so request $R$ is cut in any multicut of $P\setminus\{R\}$. 
\item[(5)] Let $x$ be an $L_2$-leaf with a wingspan $W$ dominating $k+1$ endpoint-disjoint requests. If a multicut of size $k$ exists, it contains an edge $e$ which cuts two of these requests. As the requests are endpoint-disjoint, their intersection is included in the internal tree, hence in $W$. Assume, for example, that $e$ is on the left of the leaf $x$. Then all the requests from $x$ which go to the left go through $e$, and moreover $x$ has no group request. Thus, if a solution exists, there is a solution without $e(x)$, since $e(x)$ can be replaced by the edge $e'$ which is on the right of the neighbor of $x$. \qed
\end{itemize}
\vspace{12pt}

\begin{lemma}
Deciding whether a rule applies and applying it takes polynomial time.
\end{lemma}
\proof
\par Denote by $n$ the number of nodes in $T$ and by $r$ the number of requests, which is $O(n^2)$.
\begin{itemize}
\item[(0)] The application of Rule Unit Request takes time $O(r)$. 
\item[(1)] The maximum edge-disjoint paths problem in trees is polynomial, see \cite{garg}, thus Rule Disjoint Requests is polynomial.
\item[(2)] Rule Unique Direction can be applied in time $O(rn^2)$.
\item[(3)] Rule Inclusion can be applied in time $O(r^2)$.
\item[(4)] For the running time of Rule Common Factor, consider a request $R$. Informally, we are looking for a large enough set of requests which intersect $R$, possibly leaving it at one or two places, such that the edges through which they leave are all distinct. More formally, let $Z$ be the set of edges not in $R$ but sharing a vertex with some edge in $R$. Let $Y$ be the set of edges $e$ in $Z$ such that there exists a request starting at a node in $R$ and going through $e$. We can assume without any loss that one request per such edge $e$ is chosen. Let $G$ be the graph which vertices are $Z-Y$ and which edges are the pairs $(e,e')$ such that there exists a request going through both $e$ and $e'$. There exist $k+1$ paths as in Rule Common Factor if and only if $G$ has a matching of size at least $k+1-|Y|$. As the matching problem is polynomial, the application of Rule Common Factor takes polynomial time.
\item[(5)] Let $W$ be a wingspan, let $G$ be the graph which vertices are the leaves pending from $W$ and where two leaves are adjacent if there is a request between them. There exist $k+1$ endpoint-disjoint requests dominated by $W$ if and only if $G$ has a matching of size $k+1$, thus Rule Dominating Wingspan is polynomial. \qed
\end{itemize}
\vspace{12pt}
\begin{lemma}
The reduction process has a polynomial number of iterations.
\end{lemma}
\proof
\par Each rule decreases the sum of the lengths of the requests, which is initially less than the number of requests times the number of nodes. \qed
\par In the following we consider an instance in which none of these rules can be applied, and prove that such a reduced instance has polynomial size in $k$. 

\par Let us introduce two graphs theoretic lemmas which are used in our proof.
\begin{lemma}
\label{matching}
Let $G$ be an undirected graph having $m$ edges, of maximal positive degree $\Delta$. Then $G$ has a matching of size $\lfloor{\frac{m}{2\Delta-1}}\rfloor$.
\end{lemma}
\proof
Such a matching can be obtained by a greedy algorithm, as taking an edge $uv$ in the matching forbids the edges adjacent to $u$ and those adjacent to $v$ (there are at most $2\Delta-1$ such edges, including $uv$). 
\qed

\begin{lemma}
\label{stability}
Let $H$ be an undirected graph on $n$ vertices, of maximal degree $\Delta$. Then $H$ has an independent set of size $\lfloor\frac{n}{\Delta+1}\rfloor$.
\end{lemma}
\proof
Such an independent set can be obtained by a greedy algorithm, as taking a vertex $u$ in the independent set forbids the vertices adjacent to $u$. 
\qed

\begin{theorem}
The \MIC problem has a kernel of size $O(k^5)$.
\end{theorem}
The rest of this section is dedicated to the proof of the theorem.
\begin{obs}\label{obse}
A node has at most $k+1$ R-neighbors in each direction.
\end{obs}
\proof
\par If a node $x$ has $k+2$ R-neighbors in, say, the right direction, then Rule Common Factor applies to any longest 
right request of $x$. \qed
\begin{claim}\label{bad}
There are at most $2(k+1)(2k+1)-1$ bad leaves.
\end{claim}
\proof
\par A bad leaf is connected to at most $k+1$ leaves of some given group, by Rule Common Factor. Let $G$ be the undirected graph whose vertices are the bad leaves of $T$ and where there is an edge between two leaves if there is a group request between them. The minimal degree in $G$ is at least 1, and the maximal degree is at most $k+1$. If there are at least $2(k+1)(2k+1)$ bad leaves then there are at least $(k+1)(2k+1)$ edges in $G$. Thus by Lemma~\ref{matching} there exist a matching of size $k+1$ which implies the existence of $k+1$ endpoint-disjoint (thus disjoint) group requests. In this case, Rule Disjoint Requests would apply. \qed

\begin{claim}
A wingspan has size at most $2(k+1)(4k+3)-1$.
\end{claim}
\proof
\par Let $W$ be a wingspan. As Rule Dominating Wingspan does not apply, $W$ does not dominate $k+1$ endpoint-disjoint requests. Let $W'$ be the set of leaves pending from $W$. Let $G$ be the undirected graph which vertices are the leaves in $W'$ and the nodes in $W$. For each leaf to leaf request $zy$ such that $z$ and $y$ are in $W'$, create an edge $zy$ in $G$. For each mixed-request $zy$ such that $z$ is in $W'$ and $y$ in $W$, create an edge $zy$ in $G$. Finding $k+1$ endpoint-disjoint requests is equivalent to finding a matching of size $k+1$ in $G$. The degree of a vertex $u$ in $G$ is at most $2k+2$ because there are at most $k+1$ requests in each direction for $u$ in $T$ (by Observation~\ref{obse}). Moreover, if $u$ corresponds to a node of $W'$, the degree of $u$ is at least one. Indeed, since the wingspan of $x$ is maximal, each $L_2$-leaf pending from $W$ must have a request dominated by $W$. 
\par If there are $2(k+1)(4k+3)$ $L_2$-leaves in $W'$, then $G$ contains at least $(k+1)(4k+3)$ edges, and so $G$ has a matching of size $k+1$ by Lemma~\ref{matching}, which in turn means the existence of $k+1$ endpoint-disjoint requests. \qed
\begin{claim}\label{leaves}
There are $O(k^3)$ $L_2$-leaves.
\end{claim}
\proof
\par Let $x$ be a $L_2$-leaf of wingspan $W$. By the previous claim, there are less than $2(k+1)(4k+3)$ leaves pending from $W$. At most $2(k+1)(4k+3)$ $L_2$-leaves not pending from $W$ have wingspans intersecting $W$ for each direction, as the furthest leaf (on the right) of wingspan intersecting $W$ has a wingspan which dominates all other leaves of wingspan intersecting $W$ from the right. Let $H$ be the auxillary graph on $L_2$, where two $L_2$-leaves are adjacent if their wingspans intertsect. $H$ has maximum degree less than $6(k+1)(4k+3)$ by the above discussion. By Lemma~\ref{stability}, if $T$ has at least $6(k+1)(k+2)(4k+3)$ vertices, then $H$ has a stable set of size $k+1$. Thus $T$ would have $k+1$ disjoint wingspans, and thus $k+1$ disjoint requests, a contradiction.

\begin{claim}
There are $O(k^5)$ $I_2$-nodes.
\end{claim}
\proof
\par By Claim~\ref{leaves}, there are $O(k^3)$ $I_2$-nodes with leaves. Let us bound the number of inner nodes. Let $I'$ be the set of inner nodes in $T$. Consider the graph $G$ on the set of vertices $I'$ where there is an edge $xy$ if $xy$ is a request in $T$.
\par Because of Rule Inclusion, each inner node has degree at most two in $G$ (one in each direction). Thus $G$ is a disjoint union of paths, called \emph{request paths}. The length of a request path is at most $k$ by Rule Disjoint Requests. A node with degree 1 in $G$ is an \emph{extremal inner node}.
\par Each extremal inner node must be an R-neighbor in $T$ of a leaf or of an internal node with a leaf (otherwise it would be reduced by Rule Unique Direction). Denote by $X$ the set of leaves and internal nodes with a leaf attached to it. Each node in $X$ has $O(k)$ R-neighbors among the inner nodes, and $|X|=O(k^3)$, so there are $O(k^4)$ inner nodes with a neighbor in $X$ (in particular, at most $O(k^4)$ extremal inner nodes). Each extremal inner node belongs to a unique request path of size at most $k$. Moreover each inner node with no neighbor in $X$ must belong to a request path. So there are $O(k^5)$ inner nodes in $T$. \qed
\par There are $O(k^3)$ leaves and $O(k^5)$ internal nodes in a reduced instance. Thus the \MIC problem has a kernel of size $O(k^5)$.
\qed

\section{General Trees}
\par Should no confusion arise, we retain the terminology of the previous section.
\par Let ($T$, $P$, $k$) be an instance. Let $T'$ be the tree obtained from $T$ by deleting the leaves. We partition the set of nodes  of $T$ into the following seven sets:
\begin{itemize}
\item The set $I_1$ of leaves in $T'$.
\item The set $I_2$ of degree 2 nodes in $T'$.
\item The set $I_3$ of the other nodes in $T'$. 
\item The set $L_1$ of leaves rooted at $I_1$.
\item The set $L_2$ of leaves rooted at $I_2$, endpoint of no group request.
\item The set $L_2'$ of leaves rooted at $I_2$, endpoint of at least one group request.
\item The set $L_3$ of leaves rooted at $I_3$.
\end{itemize}
We also denote by $I$ the set of internal nodes of $T$, and by $L$ the set of leaves of $T$.

\par We need a few technical definitions. 
A \emph{caterpillar} of $T$ is a maximal connected component of $T-I_3-L_3$. The \emph{backbone} of a caterpillar is the set of internal nodes of $T$ in this caterpillar. A caterpillar $C$ is \emph{non-trivial} if the set of internal nodes in $C$ seen as a caterpillar has size at least two. The \emph{extremities} of a non-trivial caterpillar $C$ are the two nodes of $C$ which are $I_2$ or $I_1$-nodes of $T$ and become $I_1$-nodes in $C$. A \emph{minimal request} of a node $x$ is a request having $x$ as an endpoint and which internal path is minimal for inclusion among all internal paths of requests with $x$ as an endpoint. If several requests have the same internal paths, we arbitrarily distinguish one as minimal and will not consider the others as minimal. If $xy$ is a minimal request of $x$ then $y$ is called a \emph{closest R-neighbor} of $x$. 
\par Let $x$ and $y$ be nodes in $T$. If $z$ lies on the path between $x$ and $y$, or is a leaf rooted at the path between $x$ and $y$, we say that $z$ lies \emph{toward} $y$ from $x$ (and we do not write "from $x$" should no confusion arise).
\par Assume $x$ is an $L_2$-leaf of a caterpillar $C$ (that is, an $L_2$-leaf of $T$ which belongs to $C$). Let $f(x)$ be the node from which $x$ is pending. Let $Gr(x)$ be the group of leaves pending from $f(x)$. Let $A(x)$ and $B(x)$ be the two connected components of $T-\{f(x)\}-Gr(x)$. Let $a(x)$ (resp. $b(x)$) be the extremity of $C$ in $A(x)$ (resp. $B(x)$). If $A(x)$ (resp. $B(x)$) contains no extremity of $C$, that is if $f(x)$ is an extremity of $C$, then we define $a(x)=f(x)$ (resp. $b(x)=f(x)$). A \emph{wingspan} $W$ of $x$ is formed by the restriction to internal nodes of the union of two requests between $x$ and two of its closest R-neighbors lying respectively in $A(x)$ and $B(x)$. Observe that $x$ can have several wingspans. The \emph{subcaterpillar} of the wingspan $W$ consists in $W$ and the leaves rooted at $W$.
\par An $L_2$-leaf $x$ \emph{covers} a caterpillar $C$ if either $x\notin C$ and there is a request starting at $x$ and going through the whole backbone of $C$, or if $x\in C$ and there are two minimal requests starting at $x$ which together cover the whole backbone of $C$.
\par We apply the following reduction rules to an instance: Rules (0), (1), (2), (3), and (4) are stated in the previous section. Rule Dominating Wingspan is split for convenience into two rules, one similar to the caterpillar case and a more general one, as follows:
\begin{itemize}
\item[(5a)] Bidimensional Dominating Wingspan: if $x$ is an $L_2$-leaf of a caterpillar $C$ with a wingspan $W$ such that $W\cap C$ dominates at least $k+1$ endpoint-disjoint requests, then we contract $e(x)$. 
\item[(5b)] Generalized Dominating Wingspan: assume that $x$ is an $L_2$-leaf of the caterpillar $C$, and that $x$ covers $C$. 
Assume that for every closest neighbor $z$ of $x$ in $A(x)$, there exist $k+1$ endpoint-disjoint requests between a node lying toward $b(x)$ from $x$ and a node toward $z$ from $a(x)$. Then we contract $e(x)$.  
\end{itemize}

Each iteration of the reduction consists in applying the first applicable rule, in the above order.

\begin{lemma}
Rules (5a) and (5b) are safe.
\end{lemma}
\proof
\par Safeness of Rule Bidimensional Dominating Wingspan follows from the safeness proof of Rule Dominating Wingspan in the previous section.
\par Assume Rule Generalized Dominating Wingspan can be applied to $x$. Let $z_1,\dots,z_l$ be the closest R-neighbors of $x$ in $A(x)$. For every $i\in\{1,\dots,l\}$, because of the $k+1$ endpoint-disjoint requests mentionned in the rule, any $k$-multicut contains an edge in the path between $z_i$ and $b(x)$. Assume that a $k$-multicut $S$ contains an edge $e''$ between $x$ and $b(x)$. Let $e'$ be the edge adjacent to $e(x)$ in the path between $x$ and $a(x)$. If $S$ contains $e(x)$, then $S-\{e(x)\}\cup\{e'\}$ is also a $k$-multicut. Indeed, any request $x,u$ with $u\in A(x)$ is cut by $e'$, and any request $x,v$ with $v\in B(x)$ is cut by $e''$. Assume now that a $k$-multicut $S$ contains no edge between $x$ and $b(x)$, then for every $i\in \{1,\dots,l\}$, $S$ must contain an edge $e_i$ in the path between $z_i$ and $f(x)$. Let $e'$ be the edge adjacent to $e(x)$ in the path between $x$ and $b(x)$. If $S$ contains $e(x)$, then $S-\{e(x)\}\cup\{e'\}$ is a $k$-multicut. Indeed, any request $x,u$ with $u\in A(x)$ is cut by an edge $e_i$, and any request $x,v$ with $v\in B(x)$ is cut by $e'$. \qed

\begin{prop}
The repeated application of these rules on the instance until none can be applied takes polynomial time.
\end{prop}
\proof 
The proof of the first five cases was made for general trees in the previous section. The polynomiality of Rule Bidimensional Dominating Wingspan follows from the proof of Rule Dominating Wingspan's polynomiality in the previous section. Deciding whether there exist $k+1$ endpoint-disjoint requests between prescribed areas can still be expressed as a matching problem as in Rule Dominating Wingspan's proof, so the application of Rule Generalized Dominating Wingspan also takes polynomial time.
\qed

\begin{theorem}
The number of nodes in a reduced instance is $O(k^6)$.
\end{theorem}

\par The rest of this section is devoted to the proof of this theorem.

\begin{claim}
$|I_1|=O(k)$
\end{claim}
\proof
\par There are at most $k$ groups of leaves with a group request, by the $k+1$ disjoint requests rule. Every group of $L_1$-leaves has a group request, otherwise any leaf of this group would be deleted by Rule Unique Direction. Every $I_1$-node has at least one $L_1$-leaf pending from it, thus $|I_1|\le k$.
\qed
\begin{claim}\label{i3}
$|I_3|=O(k)$
\end{claim}
\proof
\par In a tree, there are at most as many nodes of degree at least 3 as the number of leaves, so $|I_3|\le|I_1| \le k$.
\qed
\begin{claim}
$|L_1|=O(k^2)$ and $|L_2'|=O(k^2)$
\end{claim}
\proof
\par Each leaf in $L_1$ is a bad leaf by Rule Unique Direction, and each leaf in $L'_2$ is bad by definition. As in Claim~\ref{bad} there are at most $2(k+1)(2k+1)-1$ bad leaves in $T$. Thus $|L_1\cup L'_2|=O(k^2)$
\qed

\par We now show that: 
\begin{itemize} 
\item $|L_3|=O(k^4)$
\item $|L_2|=O(k^4)$ 
\item $|I_2|=O(k^6)$
\end{itemize}

\begin{claim}\label{grou}
The number of requests from a node $x$ to a group of leaves is at most $k+1$. 
\end{claim}
\proof
Otherwise Rule Common Factor would apply to these requests.
\qed
 
\begin{claim}
The number of requests from a node $x$ to all the $L_2$-leaves in a given caterpillar $C$ is at most $2k+2$ if $x\in C$ and $k+1$ if $x\notin C$. 
\end{claim}
\proof Otherwise there would be at least $k+2$ requests sharing the same direction between $x$ and leaves in this caterpillar, and Rule Common Factor would apply to these requests. 
\qed

\begin{claim}\label{groupgroup}
There are at most $(2k+1)(k+2)-1$ requests between two groups of leaves.
\end{claim}
\proof
Let $G$ be the bipartite graph which vertices are the leaves of the two groups $Y$ and $Z$, and where a leaf in $Y$ and a leaf in $Z$ are adjacent if there is a request between them. The maximum degree in $G$ is at most $k+1$ by Claim ~\ref{grou}, thus if there are $(2k+1)(k+2)$ requests between $Y$ and $Z$, then by Lemma~\ref{matching} there would be a matching of size $k+2$ in $G$. Thus there would be $k+2$ endpoint disjoint requests between $Y$ and $Z$, and Rule Common Factor would apply. 
\qed

\begin{claim}\label{groupcat}
The number of requests between a group of leaves $E$ and the nodes in a given caterpillar $C$ is at most $2(2k+1)(k+2)-2$.
\end{claim}
\proof
Assume by contradiction that there are at least $2(2k+1)(k+2)-1$ such requests. Let $f$ be the node in which the leaves of $E$ are rooted. If $f$ belongs to $C$, then $C-f$ has two connected components. Among these two components, we select the component $C'$ in which there is the largest number of requests from $E$. If $f$ does not belong to $C$, then we let $C'=C$. There are at least $(2k+1)(k+2)$ requests between $C'$ and $E$. Consider the undirected (bipartite) graph $G$ which vertices are the leaves of E and the nodes of $C'$, and where there is an edge between a leaf from $E$ and node from $C$ if there is a request between them. This graph has maximum degree $k+1$ by Rule Common Factor, thus by Lemma~\ref{matching}, $G$ has a matching of size $k+2$. Thus there would be $k+2$ endpoint disjoint requests, and Rule Common Factor would apply to them.
\qed

\begin{claim}\label{cat}
There are at most $2k-1$ caterpillars in $T$.
\end{claim}
\proof
There are at most $2k$ nodes in $I_1\cup I_3$. Let us call them \emph{separating nodes}. Let $r$ be one of these separating nodes. Let us consider $r$ as the root of $T$. Each caterpillar is adjacent to exactly two separating nodes. Let us associate to each caterpillar of $T$ its adjacent separating node further away from the root $r$. This mapping is a bijection, and no caterpillar is mapped on $r$, thus there are at most $2k-1$ caterpillars. 
\qed

\begin{claim}
\item $|L_3|=O(k^4)$
\end{claim}
\proof
\par We have that $|I_3|=O(k)$ by Claim~\ref{i3}. Let $X$ be an $L_3$-group rooted in $y\in I_3$. Because of Rule Disjoint Requests, at most $(2k+1)(k+1)-1$ leaves in $X$ are endpoints of group requests (by Lemma~\ref{matching} on the usual auxilliary request graph on $X$). Each leaf of $X$ must be the endpoint of at least one request, so let us count the maximal number of requests contributed by each type of nodes. By Claim~\ref{groupgroup}, and as there are at most $k$ groups of $L_1$-leaves and $k$ groups of $L_3$-leaves, at most $k((2k+1)(k+2)-1)$ leaves of $X$ have a request toward an $L_1$-leaf or an $L_3$-leaf. There are at most $2k-1$ caterpillars in $T$ by Claim~\ref{cat}, and leaves in $X$ have in total at most $2(2k+1)(k+2)-2$ R-neighbors in any caterpillar by Claim~\ref{groupcat}. Thus $O(k^3)$ leaves in $X$ are endpoints of a request toward a caterpillar node, and $I_3$ nodes can contribute for at most $O(k^2)$ requests, so $|X|=O(k^3)$. This gives $|L_3|=O(k^4)$. \qed

\begin{claim}\label{l2}
\item $|L_2|=O(k^4)$
\end{claim}

\proof
\par Assume by contradiction that $|L_2|\ge 3(2k-1)(k+1)(k+1)(4k+3)$. Let $C$ be a caterpillar of $T$ containing the maximum number of $L_2$-leaves. By Claim~\ref{cat}, there are at most $2k-1$ caterpillars in $T$, thus $C$ contains at least $3(k+1)(k+1)(4k+3)$ $L_2$-leaves. 
\par Assume first that $C$ is not covered. We obtain a contradiction as in the caterpillar case. Consider $x$ to be the $L_2$-leaf having a wingspan which intersection $\tilde{W}$ with $C$ has maximal size. Let $C'$ be the subcaterpillar of backbone $\tilde{W}$. Then $C'$ contains at least $(k+1)(4k+3)$ $L_2$-leaves, otherwise one would find $k+1$ disjoint wingspans by taking $\tilde{W}$, then a $\tilde{W_1}$ disjoint from $\tilde{W}$, then a $\tilde{W_2}$ disjoint from $\tilde{W}$ and $\tilde{W_1}$, $\dots$, and finally a $\tilde{W_k}$ disjoint from $\tilde{W}, \tilde{W_1},\dots,\tilde{W_{k-1}}$, as in Claim~\ref{leaves}. Note that the caterpillars $W,W_1,\dots,W_k$ are disjoint, as their intersections $\tilde{W}, \tilde{W_1},\dots,\tilde{W_{k}}$ with $C$ are disjoint and non-empty. Thus there would be $k+1$ disjoint requests, a contradiction. Since $\tilde{W}$ is maximal, each $L_2$-leaf $y$ in $C'$ is the endpoint of a request $r\subseteq C'$. The existence of $(k+1)(4k+3)$ such leaves means there are at least $k+1$ endpoint-disjoint requests dominated by $\tilde{W}$, by Lemma~\ref{matching} applied to the usual auxiliary request graph $G$ on the $L_2$-leaves of $C'$ (note that the maximum degree of $G$ is at most $2k+2$). Which means Rule (5a) should apply, a contradiction.
\par Assume now that $C$ is covered by some $L_2$-leaf $x$. If more than $(k+1)(4k+3)$ $L_2$-leaves in $C$ do not dominate $C$, then some wingspan of $x$ dominates $(k+1)(4k+3)$ requests, and thus dominates at least $k+1$ endpoint-disjoint requests, by the usual application of Lemma~\ref{matching}. So Rule Bidimensional Dominating Wingspan should apply, a contradiction. So at least $3(k+1)(k+1)(4k+3)-(k+1)(4k+3)$ $L_2$-leaves in $C$ cover $C$, let $X$ be the set of these leaves. Let $d_1,\dots, d_j$ be the $I_1$-nodes in $A(x)$. Note that $j\le k$. 

For such an $I_1$-node $d_i$ and a leaf $x\in X$ having at least one quasi-R-neighbor lying toward $d_i$, let us denote by $rn(x,i)$ the closest quasi-R-neighbor of $x$ toward $d_i$. Let $RN(i)$ be the set of all nodes $rn(x,i)$ for leaves $x\in X$ having at least one quasi-R-neighbor lying toward $d_i$. Note that the nodes of $RN(i)$ lie on the segment $[a(x),d_i]$. 
Denote by $x^i_1,\dots,x^i_t$ the leaves in $X$ having at least one quasi-R-neighbor lying toward $d_i$, ordered according to the distance between $a(x)$ and $rn(x,i)$, from closest to furthest. If $t\ge (k+1)(4k+3)$, denote by $X_i$ the set $\{x^i_1,\dots,x^i_{(k+1)(4k+3)}\}$.



When less than $(k+1)(4k+3)$ $L_2$-leaves in $X$ have a quasi-R-neighbor toward $d_i$, mark $d_i$ as invalid, and proceed. Note that at least one $d_i$ must be valid, as $|X|>k(k+1)(4k+3)$.
\par Now we have a list of at most $k$ sets (the sets $X_i$ for $d_i$ valid) of size $(k+1)(4k+3)$. The union $X'$ of these is of size at most $k(k+1)(4k+3) < |X|$. Thus there exists an $L_2$-leaf $z$ in $X-X'$. Consider the closest quasi-R-neighbor $n_i$ of $z$ toward a valid $d_i$. There are either $(k+1)(4k+3)$ $L_2$-leaves of $X_i$ between $z$ and $a(x)$ or $(k+1)(4k+3)$ $L_2$-leaves of $X_i$ between $z$ and $b(x)$. Thus there are $k+1$ endpoint-disjoint requests either between the subcaterpillars spanned by the segments $]z,a(x)[$ and $]a(x),n_i[$ or between the subcaterpillars spanned by the segments $]b(x),z[$ and $]a(x),n_i[$, by Lemma~\ref{matching} on the usual auxiliary request graph. In the former case Rule Common Factor applies, in the latter Rule Generalized Dominating Wingspan applies. \qed

\begin{claim}
\item $|I_2|=O(k^{6})$
\end{claim}
\proof
\par There are $O(k^4)$ internal nodes with leaves in $T$, by Claim~\ref{l2}. It remains to bound the cardinal of the set $Z$ of inner nodes in $I_2$.
\par Let $r$ be an $I_1$-node of $T$, we now consider $r$ as the root of $T$. Let $u$ be a node of $Z$. Let $C(u)$ be the caterpillar containing $u$, denote by  $a(u)$ and $b(u)$ its extremities, with $b(u)$ an ancestor of $a(u)$ with respect to $r$. Let $A(u)$ be the connected component of $T-\{u\}$ containing $a(u)$. If the node $u$ has an R-neighbor in $A(u)$, select such node $v(u)$. Note that $u$ is on the path bewteen $v(u)$ and $r$. Thus, by Rule Inclusion, $v(u) \neq v(u')$ whenever $u\neq u'$. Let $G$ be the graph with vertex set $Z$, and with edge set $\{(u,v(u))|u\in Z\}$. This graph $G$ is a disjoint union of paths. By Rule Disjoint Requests, paths in $G$ have length at most $k$. Vertices $u$ in $G$ which have no R-neighbor in $A(u)$ must be adjacent in $T$ to some node not in $Z$, by Rule Unique Direction. There are $O(k^4)$ nodes not in $Z$, each of which can have at most $k$ R-neighbors in $Z$. Indeed, a vertex cannot have two different R-neighbors in the same direction, by Rule Inclusion. Thus there are $O(k^5)$ vertices $u$ without R-neighbor in $A(u)$ in $G$, which gives that there are $O(k^6)$ vertices in $G$, which finally means that there are $O(k^6)$ inner nodes in $T$. $\square$
\par This concludes the proof of the theorem.
\qed

\section{Conclusion}
\par We have shown that the (unweighted) \MIT problem admits a polynomial kernel. This kernelization algorithm, or just some particular sequence using some of the reduction rules presented above, can be used as a preprocessing or in-processsing step in a practical algorithm.
\par This analysis might not be tight, so one can hope to improve this $O(k^6)$ bound retaining the same set of reduction rules. New reduction rules might be needed to decrease this bound even further.
\par Our technique does not seem to generalize to the weighted version of \MITB. Thus deciding whether the Weighted \MIT problem admits a polynomial kernel is still open.
\par It is not known whether the general Multicut in Graphs problem is FPT with respect to this parameter $k$, even for graphs of bounded treewidth. If it turned out to be true, then the question of the existence of a polynomial kernel for Multicut in Graphs would rise. 
\par Among the most notorious open problems on polynomial kernelization stand Directed Feedback Vertex Set and Clique Cover. Directed Feedback Vertex Set consists in deciding whether a graph admits $k$ vertices which removal makes the graph acyclic. This problem was shown to be FPT in \cite{Chen}. Clique Cover consists in deciding whether the edges of a graph can be covered by at most $k$ cliques. 

\bibliographystyle{alpha}

\end{document}